# PORÓWNANIE METOD DETEKCJI ZAJĘTOŚCI WIDMA RADIOWEGO Z WYKORZYSTANIEM UCZENIA FEDERACYJNEGO Z ORAZ BEZ WĘZŁA CENTRALNEGO
## COMPARISON OF SPECTRUM OCCUPANCY DETECTION METHODS WITH THE USE OF FEDERATION LEARNING WITH AND WITHOUT A CENTRAL NODE

Łukasz Kułacz[1]

[1] Politechnika Poznańska, Poznań, lukasz.kulacz@put.poznan.pl

**Streszczenie**: Systemy dynamicznego dostępu do widma w celu podjęcia decyzji o przydziale widma dla nowego urządzenia prze-ważnie wymagają informacji o zajętości widma, a tym samym o obecności innych użytkowników. Proste metody detekcji zajętości widma są często dalekie od niezawodnych, stąd często i z powodzeniem stosowane są algorytmy detekcji zajętości widma wspierane uczeniem maszynowym czy też sztuczną inteligencją. W celu ochrony prywatności danych użytkowników i redukcji ilości danych kontrolnych przekazywanych w systemie interesującym podejściem okazuje się użycie federacyjnego uczenia maszynowego. W tej pracy porównane zostały dwa podejścia do projektowania systemu wykorzystujące federacyjne uczenie maszynowe: z wykorzystaniem węzła centralnego oraz bez wykorzystania węzła centralnego.

**Abstract**: Dynamic spectrum access systems typically require information about the spectrum occupancy and thus the presence of other users in order to make a spectrum allocation decision for a new device. Simple methods of spectrum occupancy detection are often far from reliable, hence spectrum occupancy detection algorithms supported by machine learning or artificial intelligence are often and successfully used. To protect the privacy of user data and to reduce the amount of control data, an interesting approach is to use federated machine learning. This paper compares two approaches to system design using federated machine learning: with and without a central node.

**Słowa kluczowe**: detekcja zajętości widma, federacyjne uczenie maszynowe, sieci bezprzewodowe.

## 1. WSTĘP

Jednym z głównych wyzwań stojących przed bezprzewodowymi sieciami telekomunikacyjnymi jest efektywne wykorzystanie dostępnych zasobów częstotliwościowych. Pomiary przeprowadzone na przełomie wieków w różnych regionach świata wykazały dużą nieefektywność przydziału częstotliwości statycznej do danej usługi lub zbioru usług. Ta obserwacja była jedną z przyczyn powstania koncepcji radia kognitywnego, przedstawionej w rozprawie doktorskiej J. Mitoli III z 1999 roku, w której dostęp do widma radiowego realizowany jest dynamicznie i adaptacyjnie. W ciągu ostatnich 20 lat na całym świecie przeprowadzono liczne badania mające na celu opracowanie skutecznych metod określania zajętości widma (tzw. sensing), zarządzania widmem i elastycznego dostępu do niego. Główne wnioski z przeprowadzonych analiz to duża zawodność i niedokładność metod detekcyjnych zaimplementowanych tylko w jednym węźle sieci i wynikająca z tego konieczność stosowania kooperacyjnych rozwiązań detekcyjnych lub dodatkowych informacji zapisanych w specjalnych bazach danych. Bardzo obiecującym kierunkiem badań w kontekście wyznaczania zajętości widma jest wykorzystanie różnych metod uczenia maszynowego, zarówno w podejściu jednowęzłowym, jak i kooperacyjnym. W ostatnich latach pojawiło się wiele propozycji wykorzystania technik uczenia maszynowego w kontekście radia kognitywnego, w tym wykrywania zajętości widma. I choć rozwiązania te mogą być bardzo skuteczne, to często wiążą się z dość dużą złożonością obliczeniową, zwłaszcza w kontekście tworzenia, uczenia i aktualizowania modelu wykorzystywanego w algorytmie uczenia maszynowego. Co więcej, bardzo często algorytmy te wymagają przesyłania wielu dodatkowych danych, co pociąga za sobą pogorszenie efektywności widmowej i energetycznej całego systemu. Ponadto istotną kwestią może być zapewnienie prywatności przesyłanych danych, zwłaszcza w sytuacji, gdy w procesie detekcji będą wykorzystywane smartfony użytkowników. W tym kontekście bardzo ciekawym i popularnym rozwiązaniem jest federacyjne uczenie maszynowe, którego główną ideą jest cykliczne przekazywanie modeli zamiast bezpośrednio danych pomiarowych. Takie podejście pozwala nie tylko zapewnić ochronę prywatności danych, ale także inteligentnie uczyć się w poszczególnych węzłach, aby móc podejmować lepsze decyzje lokalnie, korzystając z wiedzy całej sieci. Algorytm ten ogranicza również ilość dodatkowych danych przesyłanych pomiędzy urządzeniami w sieci, a co więcej zapewnia znacznie większe bezpieczeństwo danych użytkownika.

W wyniku zastosowania rozproszonego i/lub scentralizowanego uczenia federacyjnego w procesie wykrywania zajętości widma radiowego można znacznie poprawić efektywność wykorzystania zasobów radiowych w sieciach telekomunikacyjnych. W tej pracy porównane zostały metody rozproszonego uczenia federacyjnego, gdzie modele są przenoszone pomiędzy sąsiednimi węzłami bez udziału węzła koordynującego, jak i wersji scentralizowanej z węzłem centralnym.

## 2. OPIS PROBLEMU BADAWCZEGO

This research was funded in whole or in part by National Science Centre (2021/41/N/ST7/01298).
For the purpose of Open Access, the author has applied a CC-BY public copyright licence to any Author Accepted Manuscript (AAM) version arising from this submission.

Zakłada się, że dla ochrony obecnych transmisji należy zapewnić wydajny algorytm wykorzystujący informacje o środowisku, uzyskiwane poprzez wykrywanie widma lub z dedykowanej bazy danych [1][6]. Szczególną uwagę zyskały dwa podejścia do elastycznego współdzielenia widma, głównie licencjonowany współdzielony dostęp (ang. *Licensed Shared Access*, LSA) i obywatelska usługa radiowa szerokopasmowa (ang. *Citizens Broadband Radio Service*, CBRS) z powiązanym systemem dostępu do widma (ang. *Spectrum Access System*, SAS). Pierwsze podejście jest bardziej popularne w Europie (wprowadzone przez Komisję Europejską [8]), drugie natomiast jest standaryzowane i wdrażane w Stanach Zjednoczonych. W tym modelu wprowadzani są dodatkowi licencjonowani użytkownicy. Mogą oni wykorzystywać fragmenty widma używane przez inne zastane systemy, o ile są zgodne z licencją na współdzielenie widma.

W wyniku prac CEPT i ETSI, określone zostały ramy współdzielenia dla LSA, które zawierają zharmonizowane warunki techniczne, koordynację transgraniczną itp. Z perspektywy architektury, wdrożenie koncepcji LSA zakłada obecność dwóch komponentów na szczycie istniejącej architektury komórkowej, głównie, repozytorium LSA i kontroler LSA. Ewolucja LSA jest również rozwijana w celu zapewnienia priorytetów użytkowników i bardziej dynamicznego ogólnego podejścia do dostępu do współdzielonych zasobów widma, niż może zapewnić system LSA opracowany dla pasma 2,3-2,4 GHz [5][7]. Drugi z najpopularniejszych schematów współdzielenia widma, CBRS, to trójwarstwowy model współdzielenia wprowadzony przez FCC w Stanach Zjednoczonych dla pasma 3550-3700 MHz [9]. W przeciwieństwie do LSA, umożliwia on dodatkowe wykorzystanie zarówno licencjonowanego, jak i nielicencjonowanego widma, o ile prawa operatora zastałego są chronione. Wprowadzane są tutaj dwa rodzaje wykorzystania widma, licencje dostępu priorytetowego (ang. *Priority Access Licensees*, PAL), czyli licencjonowani użytkownicy działający na zasadach podobnych do tych w koncepcji LSA, oraz zwolnione z licencji, czyli ogólny autoryzowany dostęp (ang. *General Authorized Access*, GAA). Dedykowany system dostępu do widma (SAS) kontroluje dostęp do widma zarówno dla użytkowników PAL, jak i GAA, jednak tylko użytkownicy PAL są chronieni przed zakłóceniami. Rozważano jednak nowe podejścia do zarządzania zakłóceniami między użytkownikami GAA, głównie CBRS Alliance omawia potrzebę wprowadzenia dedykowanego menedżera współistnienia. Schemat takiego systemu można zastosować do użytkowników tylko wtedy, gdy istnieją niezawodne metody wykrywania obecności operatora zastałego. W takim przypadku szeroko stosowane jest uczenie maszynowe.

Uczenie maszynowe jest wykorzystywane w procesie wykrywania zajętości widma zarówno w wersji jednowęzłowej, jak i wielowęzłowej [2]. Wyzwaniem jest jednak współpraca pomiędzy poszczególnymi węzłami. Model tworzony lokalnie i generowana decyzja są wystarczające, gdy pojedynczy węzeł sieci odpowiada za detekcję zajętości widma. Jeżeli kilka sąsiednich węzłów wykonuje detekcję zajętości widma, możliwa jest wymiana uzyskanych danych między tymi węzłami w celu optymalizacji podejmowanych decyzji. W jaki sposób te informacje powinny być wymieniane i jak powinny wpływać na poszczególne węzły jest istotnym problemem badawczym. Potencjalnym rozwiązaniem tego problemu jest wykorzystanie federacyjnego uczenia maszynowego. Podejmowane są pierwsze próby wykorzystania tego narzędzia w procesie detekcji zajętości widma i pojawiają się pierwsze publikacje na ten temat [3][4]. Uczenie federacyjne wydaje się być obiecującym rozwiązaniem w tym kontekście, ale nie zostało jeszcze dokładnie zbadane. W ramach tej pracy zakładamy, że dobór odpowiedniej metody uczenia maszynowego w połączeniu z odpowiednią metodą wymiany parametrów modelu powinien skutkować poprawą ogólnej skuteczności detekcji widma. Sam pomysł można zrealizować na dwa sposoby. Najpierw należy sprawdzić skuteczność różnych metod uczenia maszynowego podczas wykrywania zajętości widma w systemie rozproszonym, gdzie parametry modelu wymieniane są między sąsiednimi węzłami. Po drugie, należy sprawdzić różne metody tworzenia grup i wyboru węzłów centralnych, a dalej odpowiednie metody zbierania danych z węzłów, ich przetwarzania i doboru odpowiednich modeli dla węzłów w ramach grupy. Najprostszy koncepcyjnie algorytm federacyjnego uczenia maszynowego jest federacyjne uśrednianie (ang. *Federated Averaging*, FedAvg), który może w tym przypadku nie wystarczyć i należy przetestować inne podejścia. Ogólnym celem jest zaproponowanie algorytmu federacyjnego uczenia w wersji rozproszonej i scentralizowanej, gdzie najpierw wybierana jest odpowiednia metoda uczenia maszynowego, a następnie odpowiednia metoda wyboru i wymiany modeli. Wykorzystanie sieci neuronowych w tym kontekście wydaje się być dobrym rozwiązaniem, jednak ze względu na dużą złożoność obliczeniową może być nieakceptowalna w pojedynczych węzłach sieci.

Warto zaznaczyć, że przedstawione w pracy pomysły są innowacyjne, gdyż badania nad wykorzystaniem federacyjnego uczenia maszynowego w procesie detekcji zajętości widma są na wczesnym etapie i pokazują potencjał tego rozwiązania. Znaczenie potencjalnych wyników uzyskanych w ramach realizacji tych pomysłów to przede wszystkim poprawa skuteczności wykrywania zajętości widma w sieciach bezprzewodowych. Detekcja zajętości widma oraz nierozłączny z nią termin radia kognitywnego są od dawna przedmiotem badań. Pojawia się w normalizacji dotyczącej wykrywania widma w 3GPP. Ponadto standard CBRS w dużej mierze opiera się na wykrywaniu użytkowników pierwotnych w systemie i jest wykorzystywany komercyjnie w Stanach Zjednoczonych. W Europie, w Niemczech, jest duże zainteresowanie przemysłu wykorzystaniem pasma 3,5 GHz na podobnych zasadach jak w systemie CBRS. Z tego powodu znalezienie niezawodnych i wydajnych metod określania dostępności widma radiowego jest bardzo ważnym tematem.

## 3. DETEKCJA ZAJĘTOŚCI WIDMA

W tej sekcji przedstawione zostały dwie koncepcje opracowania algorytmu federacyjnego uczenia maszynowego do poprawy detekcji zajętości widma radiowego: wykorzystująca węzeł centralny oraz w pełni rozpro-szona. W obu przypadkach przedstawiony został opis symulacji i analizowanego systemu, który może być wykorzystany jako podstawa do weryfikacji opracowanych algorytmów.

### 3.1. Federacyjne uczenie maszynowe bez węzła centralnego

Kluczowym aspektem podczas opracowywania algorytmu federacyjnego uczenia maszynowego w celu zwiększenia efektywności wykrywania zajętości widma jest opracowanie mechanizmu wymiany i aktualizacji parametrów modelu pomiędzy najbliższymi urządzeniami w sieci. Kwestia ta dotyczy nie tylko wyboru jakie parametry wybranego modelu powinny być wymieniane między węzłami, ale również zakresu wymiany parametrów (tj. w jakim promieniu i ile urządzeń w bezpośrednim sąsiedztwie ma wymieniać informacje o modelu), a także częstotliwość ich wymiany. Dodatkowo, z punktu widzenia detekcji zajętości widma i podejmowania lokalnych decyzji, należy zaproponować sposób wykorzystania parametrów wymienionych pomiędzy węzłami (np. czy wyznaczać ich średnią wartość, czy należy ważyć parametry modelu uzyskane z różnych węzłów proporcjonalnie do odległości lub liczby próbek użytych do trenowania modelu itp.). W pierwszym kroku zalecana jest weryfikacja proponowanych mechanizmów i algorytmów za pomocą specjalnie stworzonej symulacji komputerowej. Warto rozważyć również wykorzystanie sieci neuronowych jako algorytmu odniesienia. Kluczowym jest sprawdzenie skuteczności detekcji oraz złożoności obliczeniowej poszczególnych algorytmów. Złożoność tą obliczyć można poprzez pomiar czasu obliczeń, który przekłada się na energię zużytą do przetwarzania oraz ilość pamięci potrzebnej do zbudowania modelu.

W tym celu przeprowadzenia weryfikacji przez symulację komputerową jako odniesienie warto rozważyć wygenerowanie kilku sygnałów referencyjnych za pomocą uniwersalnego radia programowalnego (ang. *Universal Software Radio Peripheral*, USRP). Za pośrednictwem innego urządzenia USRP odebrane próbki sygnału dla różnych parametrów nadawanego sygnału (modulacja sygnału, moc nadawania, długość transmisji danych, długość przerwy w transmisji itp.) powinny być zgromadzone i zapisane w plikach. Ponadto w ten sam sposób mierzony powinien być sygnał telewizyjny, którego dokładne parametry (lokalizacja nadajnika, moc nadajnika, nachylenie anteny itp.) są dostępne na stronie internetowej. Za pomocą sprzętu laboratoryjnego możliwe jest również wygenerowanie lokalnego sygnału telewizyjnego. Utworzone pliki wykorzystać można do trenowania i weryfikacji modeli uczenia maszynowego. Schemat ideowy systemu do zebrania zestawu referencyjnego próbek sygnału przedstawiono na Rys. 1.

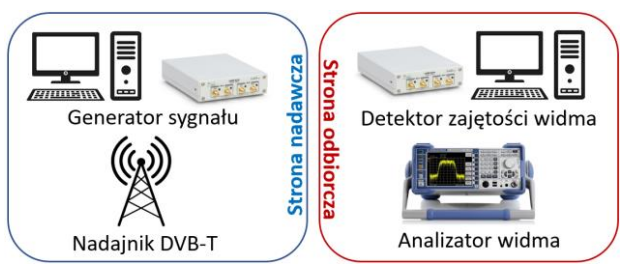

*Rys. 1. Schemat zestawu do zbierania danych referencyjnych*

Uproszczony schemat sieci, analizowanej w opisywanym przypadku przedstawiono na Rys. 2.

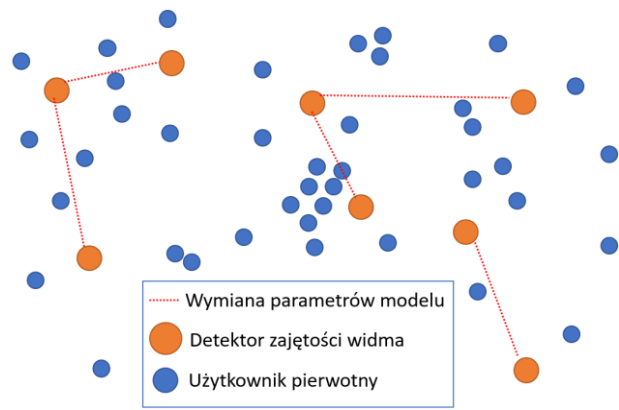

*Rys. 2. Uproszczony schemat sieci bez węzła centralnego*

W ramach symulacji na określonej powierzchni umieszczone może być kilkanaście urządzeń do wykrywania zajętości widma. Wtedy, użytkownicy pierwotni o pozycji wybranej losowo w obrębie tego samego obszaru są umieszczani w systemie wraz z modelowaniem ich transmisji, czyli czy dany użytkownik przesyła dane w danym momencie. Przedstawiona symulacja pozwoliłaby sprawdzić możliwość wykorzystania federacyjnego uczenia maszynowego do wykrywania zajętości widma w celu poprawy efektywności dynamicznego wykorzystania zasobów radiowych bez obecności węzła centralnego.

### 3.2. Federacyjne uczenie maszynowe z węzłem centralnym

Drugim rozważanym podejściem jest opracowanie algorytmu federacyjnego uczenia maszynowego z węzłem centralnym, a w szczególności opracowanie mechanizmu wymiany parametrów modelu pomiędzy urządzeniami w sieci a węzłem centralnym. Rozważany problem badawczy, ponownie, koncentruje się na określeniu częstotliwości wymiany danych między węzłami, rodzaju wymienianych danych oraz sposobu ich przetwarzania w węźle centralnym. W celu weryfikacji opracowanych mechanizmów i algorytmów, podobnie jak poprzednie, powinien zostać zaimplementowany specjalny symulator komputerowy. Warto porównać uzyskane wyniki z tej symulacji z wynikami uzyskanymi w drodze symulacji algorytmu federacyjnego uczenia maszynowego bez wykorzystania węzła centralnego. Tutaj podobnie powinna być wybrana odpowiednia metoda uczenia maszynowego, natomiast dodatkowym wyzwaniem jest zarządzanie węzłem centralnym - zbieranie i dystrybucja parametrów modelu. Podczas analizy obu algorytmów zalecane jest wykorzystanie tych samych próbek sygnału, opisanych wcześniej. Podobnie jak poprzednio, opracowany algorytm powinien zostać w drodze symulacji poprzez ocenę złożoności obliczeń, ilość danych przesyłanych do węzła centralnego z pojedynczego węzła sieci i odwrotnie. Dodatkowo istotnym aspektem jest sprawdzanie złożoności obliczeniowej algorytmu działającego w węźle centralnym. Bardzo uproszczony schemat sieci analizowanej w tym zadaniu przedstawiono na Rys. 3.

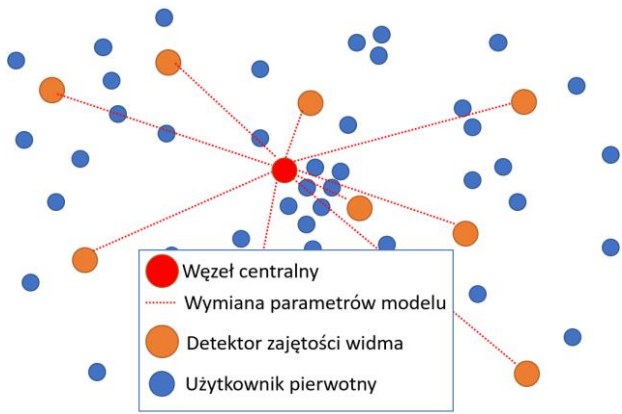

*Rys. 3. Uproszczony schemat sieci z węzłem centralnym*

W uproszczeniu sieć symulowana powinna być generowana identycznie jak w poprzedniej symulacji. Ważną różnicą jest obecność jednego dodatkowego węzła centralnego. Ma to ogromny wpływ na sposób wymiany informacji między węzłami przechodzącymi przez węzeł centralny, zamiast wymiany między sąsiednimi urządzeniami. Najbardziej złożonym aspektem w tym przypadku będzie sposób wyznaczenia współczynników modelu dla poszczególnych urządzeń. Przedstawiona symulacja pozwoliłaby na sprawdzenie możliwości wykorzystania sfederowanego uczenia maszynowego do wykrywania zajętości widma w celu poprawy efektywności dynamicznego wykorzystania zasobów radiowych w podejściu z węzłem centralnym.

Wstępna analiza rozważanych algorytmów wskazuje na kilka istotnych różnic wynikających z zastosowania węzła centralnego. Przede wszystkim węzeł centralny zwiększa ilość informacji kontrolnych przesyłanych w sieci, jednocześnie powodując duże zagęszczenie takich transmisji w okolicy węzła centralnego. Potencjalnie, może to skutkować dużym poziomem interferencji. Podsumowanie tej analizy zostało przedstawione w Tab. 1.

*Tab. 1. Porównanie algorytmów federacyjnego uczenia maszynowego*

| Aspekt | Użycie węzła centralnego | Brak węzła centralnego |
|---|---|---|
| Konieczność komunikacji z okolicznymi węzłami | opcjonalna | wymagana |
| Elastyczność względem topologii sieci | ograniczona pozycją węzła centralnego | znaczna |
| Ilość danych przesyłanych w jedno lub z jednego miejsca | znacząca; zwłaszcza w węźle centralnym | zrównoważona |
| Złożoność obliczeniowa algorytmu tworzącego wspólny model | duża; szczególnie w samym węźle centralnym | średnia; zależna od zagęszczenia węzłów |
| Spodziewana skuteczność detekcji zajętości widma | duża; ogromna ilość danych | średnia; ograniczona ilość danych |

## 4. WNIOSKI

W tej pracy przybliżono dwa podejścia do opracowania algorytmu federacyjnego uczenia maszynowego do zwiększania skuteczności detekcji zajętości widma: z użyciem i bez użycia węzła centralnego. Przedstawiona została również metoda symulacji pozwalająca na weryfikację opracowanych rozwiązań. Wstępne porównanie obu podejść może posłużyć do wyboru odpowiedniej metody do konkretnej analizowanej sieci. Wykorzystanie federacyjnego uczenia maszynowego w tym kontekście wygląda obiecująco, jednakże wymaga dalszych, obszernych badań, weryfikacji symulacyjnej oraz sprzętowej.

## 5. PODZIĘKOWANIE

## LITERATURA

[1] Astaneh Saeed, Gazor Saeed. 2013. „Relay-assisted spectrum sensing", *Communications IET*, 8 (1) : 11–18.
[2] Bkassiny Mario, Li Yang. 2013. „A survey on machine-learning techniques in cognitive radios", *IEEE Communications Surveys & Tutorials*, 15 (3) : 1136–1159.
[3] Gao Yujia, Liu Liang. 2021. „Federated sensing: edge-cloud elastic collaborative learning for intelligent sensing", *IEEE Internet of Things Journal*, 8 (14) : 11100–11111.
[4] Jeon Yo-Seb, Amiri Mohammad. 2021. „A compressive sensing approach for federated learning over massive MIMO communication systems", *IEEE Transactions on Wireless Communications*, 20 (3) : 1990–2004.
[5] Jokela Tero, Kokkinen Heikki. 2018. „Trial of spectrum sharing in 2.3GHz band for two types of PMSE equipment and mobile network", *IEEE International Symposium on Broadband Multimedia Systems and Broadcasting (BMSB),* 1–5.
[6] Kliks Adrian, Goratti Leonardo, Chen Tao. 2016. „REM: revisiting a cognitive tool for virtualized 5G networks", *23rd International Conference on Telecommunications (ICT)*, 1–5.
[7] Mustonen Miia, Matinmikko Marja. 2014. „Evaluation of recent spectrum sharing models from the regulatory point of view", *1st International Conference on 5G for Ubiquitous Connectivity*, 11–16.
[8] Radio Spectrum Policy Group. 2011. „Report on collective use of spectrum and other spectrum sharing approaches", 11–392.
[9] Sohul Munawwar, Yao Miao. 2015. „Spectrum access system for the citizen broadband radio service", *IEEE Communications Magazine*, 53 (7) : 18–25.